\documentclass[prd,preprint,superscriptaddress]{revtex4}
\usepackage{revsymb}
\usepackage{amsmath}
\usepackage{graphicx}
\usepackage{bm}
\newcommand{\be}{\begin{equation}}
\newcommand{\ee}{\end{equation}}
\newcommand{\ben}{\begin{eqnarray}}
\newcommand{\een}{\end{eqnarray}}

\begin{document}
\title{Toward a solution of the coincidence problem}
\author{Sergio del Campo\footnote{sdelcamp@ucv.cl}} \affiliation{Instituto de F\'{\i}sica,
Pontificia Universidad Cat\'{o}lica de Valpara\'{\i}so, Avenida
Brasil 2950, Casilla 4059, Valpara\'{\i}so, Chile}
\author{Ram\'{o}n Herrera\footnote{ramon.herrera@ucv.cl}}
\affiliation{Instituto de F\'{\i}sica, Pontificia Universidad
Cat\'{o}lica de Valpara\'{\i}so, Avenida Brasil 2950, Casilla
4059, Valpara\'{\i}so, Chile}
\author{Diego Pav\'{o}n\footnote{diego.pavon@uab.es}}
\affiliation{Departamento de F\'{\i}sica, Facultad de Ciencias,
Universidad Aut\'{o}noma de Barcelona, 08193 Bellaterra
(Barcelona), Spain}

\begin{abstract}
The coincidence problem of late cosmic acceleration constitutes a
serious riddle with regard to our understanding of the evolution
of the Universe. Here we argue that this problem may someday be
solved -or better understood- by expressing the Hubble expansion
rate as a function of the ratio of densities (dark matter/dark
energy) and observationally determining the said rate in terms of
the redshift.
\end{abstract}

\maketitle

\section{Introduction}
The coincidence problem of late cosmic acceleration, i.e., the
fact that the density values of dark matter, $\rho_{m}$, and dark
energy, $\rho_{\phi}$, are of the same order precisely today,
constitutes a serious challenge to our understanding of the
evolution of the Universe. The otherwise rather successful
vacuum-cold dark matter ($\Lambda$CDM) model offers no
explanation. In that model, the cosmological constant must be
fine-tuned many orders of magnitude for the ratio $r \equiv
\rho_{m}/\rho_{\phi}$ to be of order unity nowadays. This partly
explains why many authors turned to models in which the dark
energy density is no longer a constant but evolves with expansion
-see e.g. \cite{edmund} for references.

One way to alleviate the coincidence problem is to provide a
mechanism by virtue of which  $r$ tends to a constant at late
times \cite{int1, int2} or at least makes it vary slower than the
scale factor at present time \cite{softk}. This can be achieved by
introducing an interaction between dark matter and dark energy
regardless if the latter obeys the dominant energy condition  or
not \cite{int3}.

Most cosmological models assume, for the sake of simplicity, that
matter and dark energy interact only gravitationally. In the
absence of an underlying symmetry that would suppress a
matter-dark energy coupling (interaction) there is no {\em a
priori} reason to dismiss it. Further, the coupling is not only
likely but inevitable \cite{jerome} and its introduction is not
more arbitrary than assuming it to vanish.  Moreover, it has been
forcefully argued that the interaction reveals itself  when one
applies the Layzer-Irvine equation \cite{layzer} to galaxy
clusters \cite{elcio}. Likewise, the cross-correlation of galaxy
catalogs with cosmic microwave background (CMB) maps seems to
observationally favor the interaction \cite{isw}. Nevertheless, it
still remains hypothetical and more abundant, varied, and accurate
measurements are needed before its existence (or non-existence)
can be established observationally.

Cosmological models where dark matter (DM) and dark energy (DE) do
not evolve separately but interact with each other were first
introduced to justify the small value of the cosmological constant
\cite{wetterich} and currently there is a growing body of
literature on the subject -see, e.g. \cite{list} and references
therein. Further, in some holographic models of dark energy the
accelerated expansion can be traced to the interaction \cite{wd}.
Recently, various proposals at the fundamental level, including
field Lagrangians, have been advanced to account for the coupling
\cite{piazza}.

Obviously, a mechanism that makes  $r$ tend  to a constant today
or decrease its rate to a lower value than the scale factor
expansion rate ameliorates the coincidence problem significantly,
but it does not solve it in full. To do so the said mechanism must
also achieve $r_{0} \sim {\cal O}(1)$. We are not aware of any
mechanism able to fulfill the latter.

It is the view of the present authors that, on the one hand,
$r_{0}$ ought to be understood -at least, for the time being- as
an input parameter much in the same way as other key observational
quantities -say, the present value of the cosmic background
radiation temperature, the cosmological constant $H_{0}$, the age
of the Universe $t_{0}$, or the ratio between the number of
baryons and photons. And, on the other hand, its value must be
closely related to $H_{0}$ or, for that matter, to $t_{0}$.
Accordingly, if one successfully express $r$ in terms of Hubble
rate, $H$, the problem of explaining the $r_{0}$ value reduces to
the problem of explaining the present value of $H$. This is
interesting because while we are still lacking model-independent
data of $r$ at different redshifts, two sets of observational data
values -albeit scarce- of $H$ at various redshifts are now
available. The first data set was obtained by computing
differential ages of passively evolving galaxies in the redshift
range $0.1 < z < 1.8 \;$ \cite{simon}. The other data set is based
on measurements of 192 supernovae type Ia (SN Ia) and 30 radio
galaxies up to redshift $1.2$ \cite{daly1}.

Before proceeding, we wish to emphasize that, to the best of our
knowledge, no model-independent data of the evolution of the
energy densities of matter and dark energy do exist. To obtain
such evolution (e.g., with respect to redshift) one must specify
some cosmological model. It is to say, the field equations
together with the equation of state of dark energy and some
assumption about whether the components interact or not. If they
are assumed to interact, one must also provide an expression for
the interaction.

\section{Basic relations}
Let us consider  a spatially flat
Friedmann-Lema\^{\i}tre-Robertson-Walker universe dominated by a
two-component system, namely, pressureless dark matter and dark
energy, such that the components do not conserve separately but
interact with each other in a manner to be specified below. The
energy density and pressure of the dark energy, assuming it is a
quintessence field, are given by $\, \rho_\phi = %
\textstyle{1\over{2}}\dot{\phi}^2 + V(\phi)$, and $\, P_{\phi}%
= \textstyle{1\over{2}}\ \dot{\phi}^2 - V(\phi)$, respectively. If
the dark energy is a phantom field we have instead, $\, \rho_\phi
= -\textstyle{1\over{2}}\dot{\phi}^2 + V(\phi)\,$, and $\,
P_{\phi}  = -\textstyle{1\over{2}}\ \dot{\phi}^2 - V(\phi)$. The
upper-dot stands for derivative with respect to the cosmic time
and $V(\phi)$ denotes both the quintessence field potential and
phantom potential. As is usually done, we postulate that the dark
energy component (either quintessence or phantom)
obeys a barotropic equation of state, i.e., $P_{\phi} = w\,%
\rho_{\phi}$ with $w$ a negative constant of order unity and lower
than $\, -1/3 \,$ (a distinguishing feature of dark energy fields
is a high negative pressure).

We assume that the dark matter and dark energy components are
coupled through a source (loss) term (say, $Q$) whence the energy
balances take the form
\\
\begin{equation}
\dot{\rho}_{m} + 3 H \rho_{m}  = Q \, , \qquad {\rm and} \qquad
\dot{\rho}_{\phi} + 3 H (\rho_{\phi} + P_{\phi}) = -Q \, .
\label{cons}
\end{equation}
\\
In what follows we shall consider $Q >0$. On the one hand, this
choice ensures that $r$ decreases monotonously with cosmic
expansion (as expected in usual models of cosmic structure
formation) and that around present time it varies very slowly,
i.e., slower than the scale factor. On the other hand, it shows
compatibility with the second law of thermodynamics
\cite{secondl}. We note, parenthetically, that if $Q$ were
negative (i.e., if dark matter decayed into dark energy), it would
exacerbate the coincidence problem since the ratio $r$ would
decrease faster than in the $\Lambda$CDM model -see Eq. (\ref{rr})
below.

In virtue of the above expressions and Friedmann's equation,
\\
\begin{equation}
3 H^2 = \kappa^{2} (\rho_{m} + \rho_{\phi}) \quad
\qquad(\kappa^{2} \equiv 8\pi\, G)\, ,
\label{Fried1}
\end{equation}
\\
the time evolution of $r$ can be written as
\\
\begin{equation}
\dot{r} = 3 H
r\left[w+\frac{\kappa^{2}\,Q}{9\,H^3}\frac{(r+1)^2}{r} \right] \,
. \label{rr}
\end{equation}

Using the relationship $\dot{r}=\dot{H}\,dr/dH \,$, where
\\
\[
\dot{H} = -\frac{\kappa^{2}}{2}\, (\rho_{m} + \rho_{\phi} +
P_{\phi}) = -\frac{3}{2}\; \frac{1+w+r}{1+r}\; H^{2} \, ,
\]
\\
Eq. (\ref{rr}) becomes
\\
\begin{equation}
\frac{dr}{dH} = \frac{\Im}{H} \, .
\label{rrH}
\end{equation}
\\
Here
\\
\begin{equation}
\Im=-2\,r\,\frac{(1+r)}{(1+w+r)}\left[w+\frac{\kappa^{2}\,Q}{9\,H^3}\frac{(r+1)^2}{r}
\right]\, .
\label{F}
\end{equation}
\\
Clearly, Eq.(\ref{rrH}) can be integrated whenever an expression
for $Q$ in terms of $H$ and $r$ is given.

Notice that to have $dr/dH>0$ (a very reasonable assumption since
one expects that $H(t<t_{0}) > H_{0}$) the relationship
\\
\begin{equation}
|w|>\frac{\kappa^{2}\,Q}{9\,H^3}\frac{(r+1)^2}{r} \, \label{cond}
\end{equation}
\\
must be fulfilled.

We note in passing that for the $\Lambda$CDM model (i.e., $w = -1
\, $ and $Q = 0$) one follows $\, H = %
H_{0} \,[(1+r)/(1+r_{0})]^{1/2}$.

\section{Two choices for $Q$}
In the following, we focus on two phenomenological expressions for
$Q$ previously considered in the literature. As we have carefully
checked, both models fit very well the observational data of Daly
{\it et al.} obtained from selected sets of SN Ia and $30$ radio
galaxies \cite{daly1}. In particular, the maximum variation of
$P_{\phi}$ does not exceed $23\%$ in the redshift interval $0 \leq
z \leq 1$, thus showing compatibility with the data analysis of
Daly {\it et al.} \cite{daly1}.

\noindent $(i)\,$ $Q=3\,\alpha\,H\,(\rho_\phi+\rho_m)\, $ with
$\alpha$ a positive-definite, dimensionless constant.

\begin{figure}[th]
\includegraphics[width=5.5in,angle=0,clip=true]{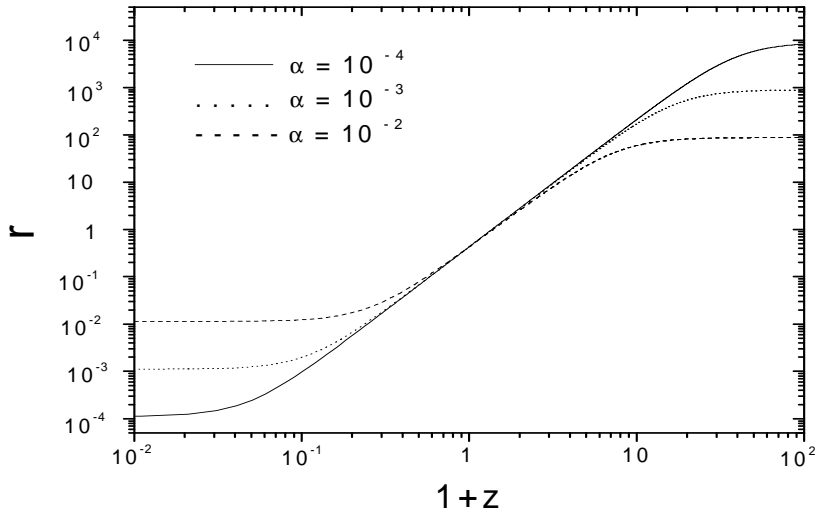}
\caption{Evolution of the ratio $r=\rho_{m}/\rho_{\phi}$ with
redshift for different values of the parameter $\alpha$. In
drawing the curves we have fixed $r_{0}= 3/7 \,$ and $\, w= -0.9$.
} \label{fig:rvsz1}
\end{figure}

This choice has been studied in detail
\cite{int1}-\cite{int3},\cite{isw, gfd}. As depicted in
Fig.\ref{fig:rvsz1}, it leads to a constant but unstable ratio $r$
at early times (high redshifts) and a lower, also constant but
stable (attractor) ratio at late times. When the dark energy is of
phantom type (not shown in the figure), the evolution of $r$ is
also similar. Further, it fits rather well data from SN Ia, cosmic
microwave background radiation (Wilkinson Microwave Anisotropy
Probe 3 yr), and structure formation -namely, Sloan Digital Sky
Survey, and Two Degree Field Redshift Survey- provided $\alpha <
2.3 \times 10^{-3}$ \cite{gfd}. Likewise, it shows compatibility
with cross correlations of galaxy catalogs with CMB maps provided
$\alpha < 0.1$ \cite{isw}.
In this case Eq.(\ref{F}) reduces to
\begin{equation}
\Im = -2 \,r\; \frac{1+r}{1+w+r}\; \left[w+\alpha\frac{(r+1)^2}{r}
\right] \, , \label{F1}
\end{equation}
where we used Eq. (\ref{Fried1}). The condition given by Eq.
(\ref{cond}) leads to restriction $|w|>\alpha\,(1+r)^2/r$.

Using last equation in (\ref{rrH}) and integrating, the Hubble
function is found to be
\\
\begin{equation}
H=H_{0}\,\exp{[I(r)-I(r_{0})]} \label{HH},
\end{equation}
\\
where $I(r)$ stands for the real part of $\tilde{I}(r)$ with
\\
\begin{equation}
\tilde{I}(r)=\frac{1}{2}\ln(1+r)-\frac{1}{4}\ln[-w\, r
-\alpha(1+r)^2]-\, \frac{1}{2}\frac{w+2}{\sqrt{-w(4\alpha + w)}}
\;\tan^{-1}\left[\frac{2\alpha(1+r)+ w}{\sqrt{-w \,(4\alpha \,
w)}}\right]\, . \label{j2}
\end{equation}

\begin{figure}[th]
\includegraphics[width=3.5in,angle=0,clip=true]{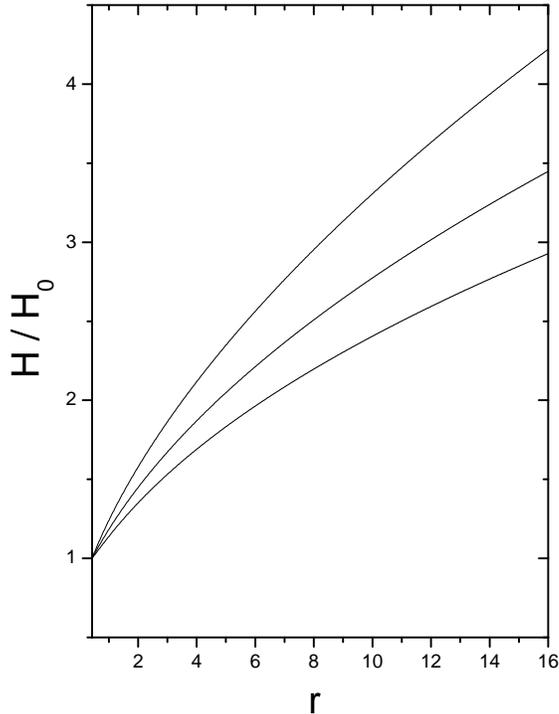}
\caption{Evolution $H$ vs the ratio $r=\rho_{m}/\rho_{\phi}$ as
given by Eq.(\ref{HH}) with $\alpha = 10^{-4}$ for $w= -0.9$
(quintessence, top line) and $w= -1.1$ (phantom, bottom line).
Also shown is the prediction of the $\Lambda$CDM model (middle
line). In drawing all the curves we have fixed $\, r_{0} = 3/7$.}
\label{fig:ratio1l1}
\end{figure}

Figure \ref{fig:ratio1l1} shows the dependence of the Hubble
function on the densities ratio for $\alpha = 10^{-4}$ and two
values of the equation of state parameter, $w$. Likewise, we have
plotted the prediction of the $\Lambda$CDM model. In all the cases
$r_{0} = 3/7$. For other $\alpha$ values compatible with the
restriction $\alpha < 2.3 \times 10^{-3}$ set by the Wilkinson
Microwave Anisotropy Probe 3 yr experiment the corresponding
graphs (not shown) are rather similar.
\\

Unfortunately, as said above, no model-independent  data of $r$ at
different values of $H$ (or $z$) exists whereby, for the time
being, we cannot directly contrast this model with observation.
This is why we turn to determine the dependence of the Hubble
factor with redshift and compare it with the two available
observational data sets $H$ vs $z$. The first model-independent
data set was  obtained from the study of the differential ages of
32 -carefully selected- passively evolving galaxies in the
redshift range $0.1 < z < 1.8 \;$ \cite{simon}. The age of each
galaxy was found by constraining the age of its older stars with
the use of synthetic stellar population models. The differential
ages roughly yields $dz/dt$, then $H(z)$ is given by   $H =
-dz/[(1+z)\,dt]\,$ -see right panel of Fig. 1 in Ref.
\cite{simon}. The second model-independent data set (see lower
panels in Fig. 9 of Daly et al. \cite{daly1}) was obtained by
applying the  model-independent analysis method of Ref.
\cite{djorgovski} to the coordinate distances of $192$  SN Ia of
Davis {\it et al.} \cite{davis} and $30$ radio galaxies of Ref.
\cite{daly2}.

To obtain the expression for H(z) of this model we combine Eq.
(\ref{HH}) with the integral of (\ref{rr}) in terms of $z$ which
is
\\
\begin{equation}
r(z) =  {\rm Re} \left\{\frac{1}{2\alpha}\left[-w+\gamma \tan
\left( A -\frac{3}{2} \, \gamma \, \ln(1+z)\right) \right]
\right\} -1 \, , \label{rz1}
\end{equation}
\\
where ${\rm Re} \,$  specifies the real part of the corresponding
quantity, $\, A = \tan^{-1} [(w + 2\alpha (1+r_{0}))/\gamma]\,$,
and $\; \gamma = \sqrt{- w \, (w + 4 \, \alpha)}$.

\begin{figure}[th]
\includegraphics[width=4.5in,angle=0,clip=true]{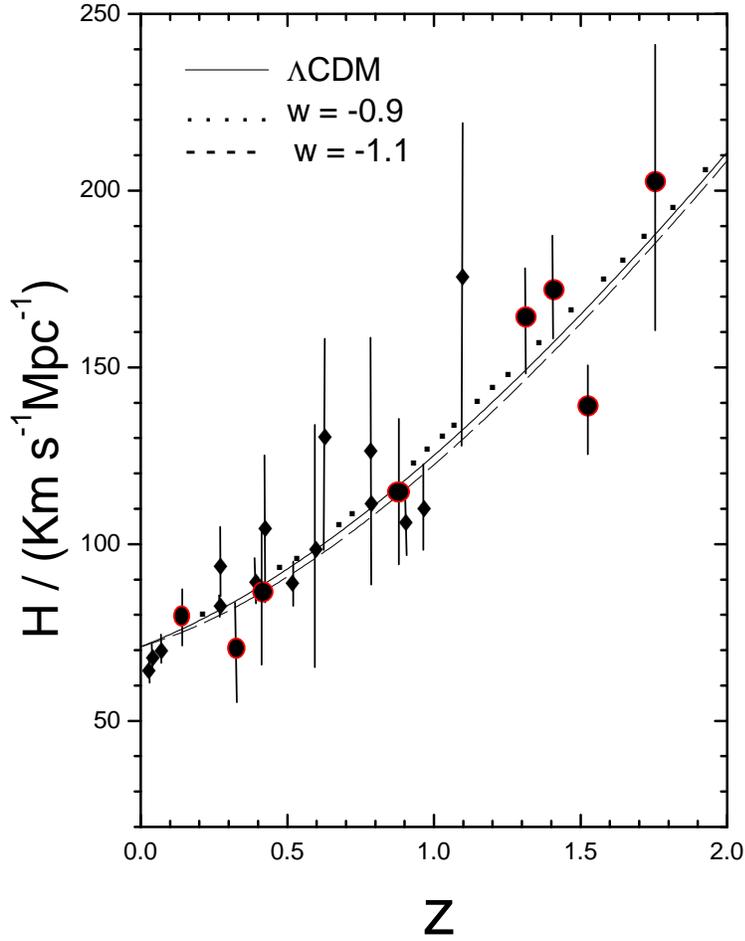}
\caption{Evolution $H$ vs $z$ with $\alpha = 10^{-4}$ for $w=
-0.9$ (quintessence) and $w= -1.1$ (phantom). Also shown is the
prediction of the $\Lambda$CDM model (solid line). In all the
cases we have fixed $r_{0} = 3/7\, $ and $H_{0} = 71 \, {\rm
km/s/Mpc}$. The data points with their $1\sigma$ error bars are
borrowed from Simon, Verde, and Jim\'{e}nez, Ref. \cite{simon}
(full circles) and table 2 of Ref. \cite{daly1} (full diamonds).}
\label{fig:hz1}
\end{figure}

\begin{figure}[th]
\includegraphics[width=6.5in,angle=0,clip=true]{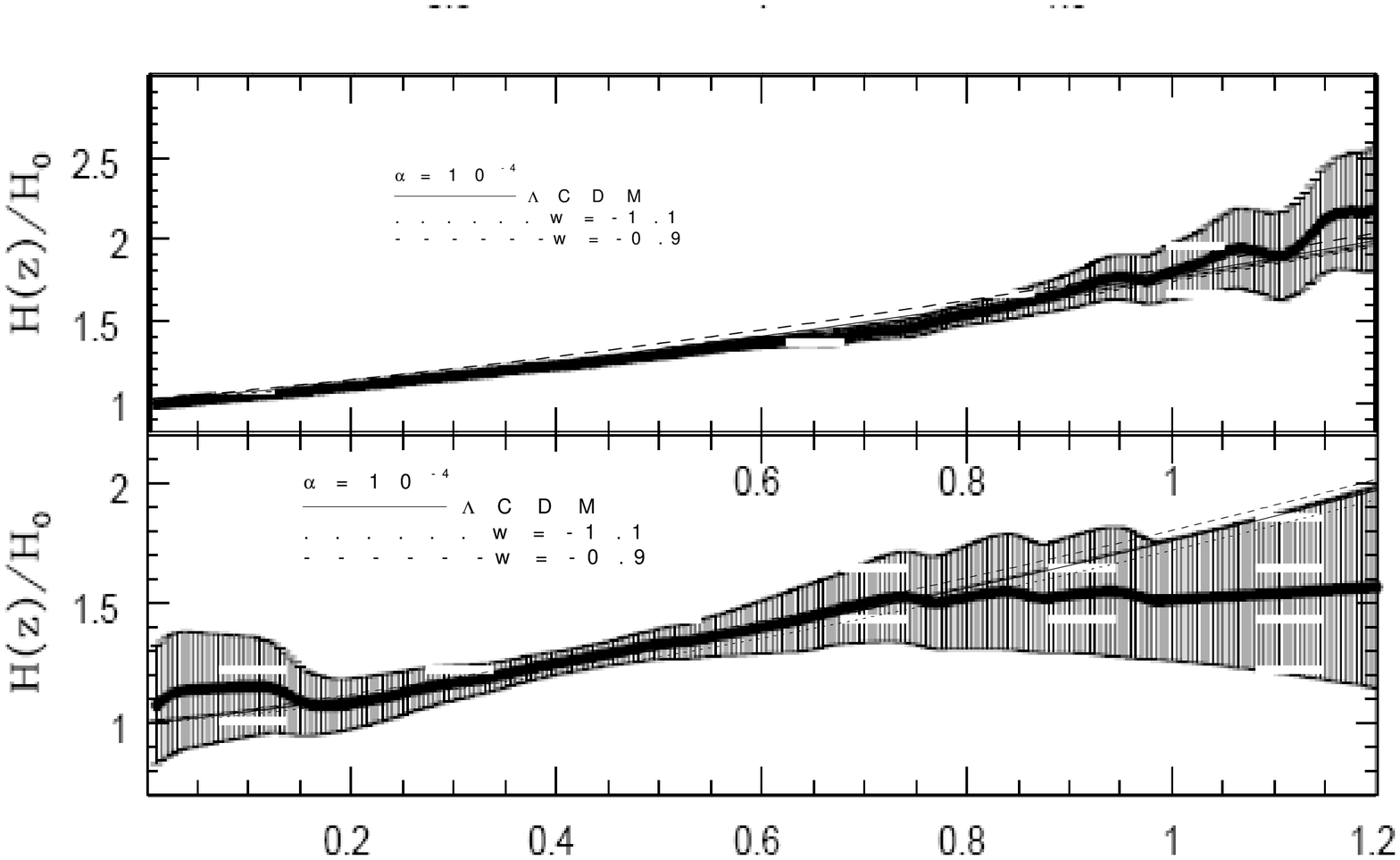}
\vspace*{-3cm}\caption{Same as Fig. \ref{fig:hz1}. The upper panel
shows the combined sample of $132$ SN Ia of Ref. \cite{davis} and
$30$ radio galaxies of Ref. \cite{daly2}. The bottom panel shows
the $30$ radio galaxies only. The data points with their $1\sigma$
error intervals and the best-fit curve (big solid line) in both
panels are borrowed from Daly {\it et al.}, Ref. \cite{daly1}.}
\label{fig:hz2}
\end{figure}

Figures \ref{fig:hz1} and \ref{fig:hz2} show the dependence of the
Hubble expansion rate on redshift predicted by the model for
$\alpha = 10^{-4}$ and two values of the equation of state
parameter, $w$. For comparison, we have also plotted the
prediction of the $\Lambda$CDM model. The observational data in
Fig. \ref{fig:hz1} are taken from Simon, Verde, and Jim\'{e}nez
\cite{simon} (full circles) and Daly {\it et al.} (full diamonds),
and in Fig. \ref{fig:hz2} from Daly {\it et al.} (cf. lower panels
of figure 9 in Ref. \cite{daly1}). As can be seen in both figures,
the fit to data does not vary significantly between graphs. Given
the scarcity of data  it is not worthwhile to compare the
$\chi^{2}$ of the different curves.

\noindent $(ii)$ $Q=3\,\beta\,H\,\rho_{m} \,$ with $\beta$ a
dimensionless positive-definite small constant.

\begin{figure}[th]
\includegraphics[width=4.0in,angle=0,clip=true]{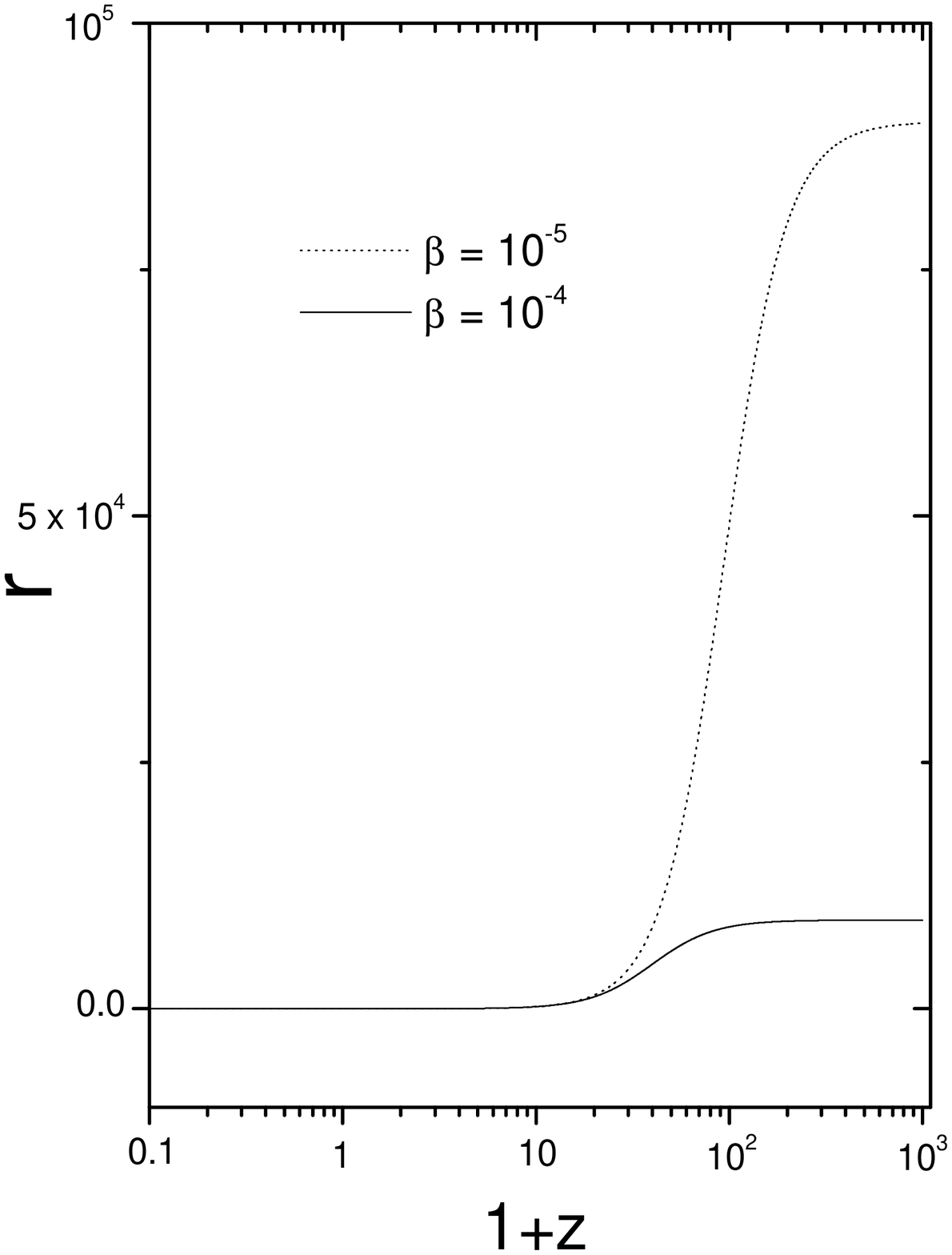}
\caption{Evolution of the ratio $r=\rho_{m}/\rho_{\phi}$ with
redshift for different values of the parameter $\beta$. In drawing
both curves we have fixed $r_{0}= 3/7 \,$ and $\, w= -0.9$. As it
is apparent, $\dot{r}/r \simeq 0$ at late times.}
\label{fig:rvsz2}
\end{figure}

Upon this choice, considered in Ref. \cite{softk}, the ratio $r$
does not tend to a constant (i.e., unlike the previous case no
attractor exists \cite{wdl}) but, as Fig. \ref{fig:rvsz2} shows,
it varies very slowly at late times -by very  slowly we mean that
$\mid(\dot{r}/r)_{0}\mid \lesssim H_{0}\, $, ``soft coincidence"-
whereby the coincidence problem gets also significantly
alleviated.
\\

Likewise, in this instance, Eq.(\ref{F}) reduces to
\\
\begin{equation}
\Im = -2\,r \;\frac{1+r}{1+w+r}\, \left[w+\beta\;(r+1) \right]\, ,
\label{F21}
\end{equation}
\\
where we used Eq.(\ref{Fried1}) to obtain
$Q=9\beta\,r\,H^3/[\kappa^{2}\,(1+r)]$. Now, the condition set by
Eq.(\ref{cond}) boils down to $|w| >\beta\,(1+r)$.

With the help of  Eqs.(\ref{rrH}) and (\ref{F21}) the Hubble
function can be cast as
\\
\begin{equation}
H = H_{0}\,\frac{I(r)}{I(r_{0})} \, , \label{HH2}
\end{equation}
\\
where
\\
\begin{equation}
I(r)=\sqrt{(1+r)}\;\;r^{[-w-1]/[2(\beta + w)]}\; [\beta (1+r)+
w]^{(1-\beta)/[2(\beta + w)]} \; . \label{j2m}
\end{equation}

Numerically, the dependence of the Hubble expansion rate upon $r$
is very close to the previous case $(i)$ for $\beta$ values
similar to $\alpha$. This is why we do not show it here.

The next step is to compute the dependence of the Hubble expansion
rate on redshift. This can be done by combining Eq. (\ref{HH2})
with
\\
\begin{equation}
r(z) = \frac{\xi \,r_{0}}{\beta\,r_0-(1+z)^{-3\xi} \,
[\beta(1+r_0)+ w]} \, , \label{rfz}
\end{equation}
\\
where $\xi= -w -\beta>0$. The latter expression follows from
integrating Eq. (\ref{rr}) in terms of $z$.

Again, the corresponding figures for $H(z)$ is rather similar to
Figs. \ref{fig:hz1} and \ref{fig:hz2} whereby we do not depict
them here.

\section{Discussion \label{discuss}}
The current value of the ratio $r = \rho_{m}/\rho_{\phi} \,$ is of
order unity (more precisely, about $3/7 \,$ -see e.g.
\cite{spergel}). This represents the ``coincidence problem" that
the $\Lambda$CDM model seems unable to account for. Nowadays,
$r_{0}$ cannot be derived from first principles and like some
other key cosmic quantities it has to be considered an input
parameter. It seems obvious, nevertheless, that in Einstein
relativity it must be closely tied to the present value of the
Hubble expansion rate, $H_{0}$, another quantity whose value we
cannot explain but that contrary to $r_{0}$  it does not mean any
puzzle. Therefore, a possible avenue to someday solve the
coincidence problem may well be to express $r$ as a function of
$H$ (or vice versa) in models such that $r$ at late times either
tends to constant or varies slowly enough (i.e., slower than in
the $\Lambda$CDM model). Thus, the problem of explaining $r_{0}$
reduces to the less acute problem of explaining $H_{0}$.

In this paper we have taken a first step in that direction.
Specifically, we have considered two phenomenological models,
previously studied in the literature, in which dark energy and
dark matter do not evolve separately but interact with each other
in such a way that either $r$ stays constant at late times (case
$(i)$) or varies very slowly (case $(ii)$) and, in both instances,
we have related $H$ to $r$. Since model-independent data relating
these two quantities are not currently available we have expressed
$r$ in terms of $z$ [Eqs. (\ref{rz1}) and (\ref{rfz}),
respectively] and resorted to the observational data of Refs.
\cite{simon} and \cite{daly1} that link $H$ to $z$. Regrettably,
given the scarcity and dispersion of the data (cf. Figs.
\ref{fig:hz1} and \ref{fig:hz2}) they are unable to discriminate
between models .

Nevertheless, it is fair to say  that in order not to spoil the
well established primeval nucleosynthesis scenario,
$\Omega_{\phi}$ should not exceed $5\%$ at that early epoch -see
Bean, Hansen, and Melchiorri  \cite{rbean}. At any rate, as Figs.
\ref{fig:rvsz1} and \ref{fig:rvsz2} show, the models considered in
this work are consistent with this constraint which, in any case,
was determined using specific non-interacting dark energy models.

It is to be hoped that in the not far future abundant and accurate
observational data, in an extended redshift range, may help
determine $H$ in terms of $z$  and  solve (or, at least, help to
better understand) the coincidence problem of late acceleration.
Yet, it may well be that the $\Lambda$CDM model comes to fit the
data better than any interacting model. In such a case, we should
conclude that the coincidence problem is just that: a coincidence
and, therefore, no longer a problem.

\acknowledgments{S.d.C. was supported from Comisi\'{o}n Nacional
de Ciencias y Tecnolog\'{\i}a (Chile) through FONDECYT Grants No.
1040624, No. 1051086, and No. 1070306 and by the PUCV. R.H. was
supported by the ``Programa Bicentenario de Ciencia y
Tecnolog\'{\i}a" through the Grant ``Inserci\'on de Investigadores
Postdoctorales en la Academia" \mbox {N$^0$ PSD/06}. D.P.
acknowledges ``FONDECYT-Concurso incentivo a la cooperaci\'{o}n
internacional" No. 7070003, and is grateful to the ``Instituto de
F\'{\i}sica" for warm hospitality; also D.P. research was
partially supported by the ``Ministerio Espa\~{n}ol de
Educaci\'{o}n y Ciencia" under Grant No. FIS2006-12296-C02-01.}


\end{document}